\documentclass{llncs}
\usepackage{latexsym,amssymb,amsmath}
\usepackage{cite}
\newcommand{\onemat}[0]{{\mathbf 1}}
\newcommand{\diag}{{\mathop{\textup{diag}}\nolimits}}

\newcommand{\GR}[0]{\textup{GR}}

\newcommand{\C}[0]{{\mathbb{C}}}
\newcommand{\F}[0]{{\mathbb{F}}}
\newcommand{\N}[0]{{\mathbb{N}}}

\newcommand{\Z}[0]{{\mathbb{Z}}}

\newcommand{\nix}[1]{}

\def\ket#1{\left|#1\right>}

\def\qed{\quad{$\Box$}}

\def\tr{\mathop{{\rm tr}}\nolimits}

\newcommand{\scal}[2]{\langle #1|#2\rangle}

\begin{document}
\title{Constructions of Mutually Unbiased Bases}
\author{Andreas Klappenecker\inst{1} and Martin R\"otteler\inst{2}}
\institute{
Department of Computer Science,\\
Texas A\&M University, College Station, TX 77843-3112, USA\\
\email{klappi\symbol{64}cs.tamu.edu}\\
\and
Department of Combinatorics and Optimization,\\
University of Waterloo, Waterloo, Ontario, Canada, N2L 3G1\\
\email{mroetteler\symbol{64}math.uwaterloo.ca}
}
%\begin{center}
%{\Large \bf On Mutually Unbiased Bases}
%\end{center}

\maketitle

\begin{abstract}
\noindent 
Two orthonormal bases $B$ and $B'$ of a $d$-dimensional complex
inner-product space are called mutually unbiased if and only if
$|\scal{b}{b'}|^2=1/d$ holds for all $b\in B$ and $b'\in B'$.  The
size of any set containing pairwise mutually unbiased bases of $\C^d$
cannot exceed $d+1$.  If $d$ is a power of a prime, then extremal sets
containing $d+1$ mutually unbiased bases are known to exist.  We give
a simplified proof of this fact based on the estimation of exponential
sums. We discuss conjectures and open problems concerning the maximal
number of mutually unbiased bases for arbitrary dimensions.
\medskip

\keywordname\ Quantum cryptography, quantum state estimation, 
Weil sums, finite fields, Galois rings.
\end{abstract}

%%%%%%%%%%%%%%%%%%%%%%%%%%%%%%%%%%%%%%%%%%%%%%%%%%%%%%%%%%%%
%
% Section: Introduction
%
%%%%%%%%%%%%%%%%%%%%%%%%%%%%%%%%%%%%%%%%%%%%%%%%%%%%%%%%%%%%

\section{Motivation}
The notion of mutually unbiased bases emerged in the literature of
quantum mechanics in 1960 in the works of
Schwinger~\cite{schwinger60}.  Two orthonormal bases $B$ and $B'$ of
the vector space~$\C^d$ are called \textsl{mutually unbiased}\/ if and
only if $|\scal{\,b\,}{\,b'\,}|^2=1/d$ holds for all $b\in B$ and all
$b'\in B'$.  Schwinger realized that no information can be retrieved
when a quantum system which is prepared in a basis state from
$B'$ is measured with respect to the basis~$B$. A striking application
is the protocol by Bennett and Brassard~\cite{bennett84} which
exploits this observation to distribute secret keys over a public
channel in an information-theoretically secure way (see
also~\cite{BT:2000}).

Any collection of pairwise mutually unbiased bases of $\C^d$ has
cardinality $d+1$ or less,
see~\cite{bandyopadhyay02,delsarte75,hoggar82,kabatiansky78,wootters89}.
Extremal sets attaining this bound are of considerable
interest. Ivanovi\'c showed that the density matrix of an ensemble of
$d$-dimensional quantum systems can be completely reconstructed from
the statistics of measurements with respect to $d+1$ mutually unbiased
bases~\cite{ivanovic81}.  Furthermore, he showed that the density
matrix cannot be reconstructed from the statistics of fewer
measurements.

Let $N(d)$ denote the maximum cardinality of any set containing
pairwise mutually unbiased bases of $\C^d$. It is known that
$N(d)=d+1$ holds when $d$ is a prime power,
see~\cite{ivanovic81,wootters89,bandyopadhyay02}. We derive a
simplified proof of this result, which takes advantage of Weil-type
exponential sums. We present two different constructions---both based on Weil sums over finite fields---in the case of odd prime
power dimensions. We exploit exponential sums
over Galois rings in the case of even prime power dimensions. 
If the dimension $d$ is not a prime power, then the exact value 
of $N(d)$ is not known.  We discuss lower bounds, conjectures, and 
open problems in the fourth section.

%%%%%%%%%%%%%%%%%%%%%%%%%%%%%%%%%%%%%%%%%%%%%%%%%%%%%%%%%%%%
%
% Section: A direct construction for prime power dimension
%
%%%%%%%%%%%%%%%%%%%%%%%%%%%%%%%%%%%%%%%%%%%%%%%%%%%%%%%%%%%%

\section{Odd Prime Powers}\label{odd}

Let $\F_q$ be a finite field with $q$ elements which has odd characteristic
$p$. Denote the absolute trace from $\F_q$ to the prime field $\F_p$
by $\tr(\,\cdot\,)$.  Each nonzero element $x\in \F_q$ defines a
non-trivial additive character $\F_q\rightarrow \C^\times$ by
$$y \mapsto\omega_p^{{\rm tr}(x y)},$$ where $\omega_p=\exp(2\pi i/p)$
is a primitive $p$-th root of unity. 
All non-trivial additive characters are of this form.

\nix{
Recall that two different kinds of characters can be attached to a
finite field $\F_q$, where $q=p^r$ for a prime number $p$:
characters of the {\em additive} group, which is isomorphic to
$(\Z_p)^r$ as well as characters of the multiplicative group, which is
isomorphic to $\Z_{p^r-1}$. In the following we use the additive
characters only and briefly recall their definition: let ${\rm tr}$
denote the trace map from $\F_q$ to the prime field $\F_p \subseteq
\F_q$ and let $\omega_p$ be a primitive $p$th root of unity in the
field $\C$ of complex numbers. Then for each $x \in \F_q$ we have an
additive character defined by $y \mapsto \omega_p^{{\rm tr}(x
y)}$. Because of the linearity of the trace it is straightforward to
see that this indeed defines a homomorphism from the additive group of
$\F_q$ to $\C$. Conversely, each additive character of $\F_q$ is of
this form. An additive character is called non-trivial if the
corresponding element $x\in \F_q$ is non-zero.
}
\begin{lemma}[Weil sums]\label{weil}
Let\/ $\F_q$ be a finite field of odd characteristic and 
$\chi$ a non-trivial additive character of\/
$\F_q$. Let $p(X)\in \F_q[X]$ be a polynomial of degree~2.
Then 
\[ 
\left|\sum_{x \in \F_q} \chi(p(x)) \right| = \sqrt{q}.
\]
\end{lemma}
We refer to \cite[Theorem 5.37]{LN:94} or \cite[p.~313]{bollobas85}
for a proof. We will use this lemma in the following constructions of
mutually unbiased bases.
\medskip

\textit{Convention.} In the following, we will tacitly assume that
the elements of $\F_q$ are listed in some fixed order, and 
this order will be used whenever an object indexed by elements of 
$\F_q$ appears.\medskip

We begin with a historical curiosity.  Schwinger introduced the
concept of mutually unbiased bases in 1960.  However, he did not
construct extremal sets of mutually unbiased bases, except in low
dimensions, and no further progress was made during the next twenty
years. Alltop constructed in 1980 complex sequences with low
correlation for spread spectrum radar and communication
applications~\cite{Alltop:80}. It turns out that the sequences given
by Alltop provide $p+1$ mutually unbiased bases in dimension~$p$, for
all primes $p\ge 5$. Unfortunately, Alltop was not aware of his
contribution to quantum physics, and his work was not noticed until
recently.  Our first construction generalizes the Alltop sequences to
prime power dimensions.

\begin{theorem}
Let $\F_q$ be a finite field of characteristic $p\ge 5$. 
Let $B_\alpha$ denote the set of vectors 
$$ B_\alpha = \{ b_{\lambda,\alpha}\,|\,
\lambda \in \F_q\},\qquad
b_{\lambda,\alpha} = \frac{1}{\sqrt{q}} 
\left(\omega_p^{\tr((k+\alpha)^3 + \lambda (k+\alpha))}
\right)_{k\in \F_q}.
$$ 
The standard basis and the sets $B_\alpha$, with $\alpha\in \F_q$, form
an extremal set of $q+1$ mutually unbiased bases of the vector space\/ $\C^q$.
\end{theorem}
\begin{proof}
Notice that $B_\alpha$ is an orthonormal basis because 
$$ 
\scal{b_{\kappa,\alpha}}{b_{\lambda,\alpha}} = 
\frac{1}{q}\sum_{k\in \F_q} 
\omega_p^{\tr((\kappa-\lambda)(k+\alpha))}. 
$$ 
Indeed, the right hand side equals 0 when $\kappa \neq \lambda$
because the argument $k+\alpha$ ranges through all values of $\F_q$; 
and equals $1$ when $\kappa=\lambda$.  

Note that all components of the sequence $b_{\lambda,\alpha}$ have
absolute value $1/\sqrt{q}$, hence the basis $B_\alpha$ and the
standard basis are mutually unbiased, for any $\alpha \in \F_q$. 

By computing the inner product $| \langle b_{\kappa, \alpha},
b_{\lambda, \beta} \rangle |$ for $\alpha \neq\beta$, we see that the
terms cubic in $k$ cancel out and, moreover, that the exponent is
given by the trace of a quadratic polynomial in $k$. By Lemma
\ref{weil} the inner product evaluates to $q^{-1/2}$, hence
$B_\alpha$ and $B_\beta$ are mutually unbiased.\qed
\end{proof}

\begin{remark}
A remarkable feature of the previous construction is that knowledge of
one basis $B_\alpha$ is sufficient because shifting the indices by
adding a field element yields the other bases. The construction does
not work in characteristic~2 and 3 because in these cases the sets
$B_\alpha$ and $B_\beta$, with $\alpha\neq \beta$, are not mutually
unbiased.
\end{remark}

Ivanovi\'c gave a fresh impetus to the field in 1981 with his seminal
paper~\cite{ivanovic81}. Among other things, he gave explicit
constructions of $p+1$ mutually unbiased bases of $\C^p$, for $p$ a
prime.  His construction was later generalized in the influential
paper by Wootters and Fields~\cite{wootters89}, who gave the first
proof of the following theorem.  This proof was recently rephrased by
Chaturvedi~\cite{chaturvedi02}, and an alternate proof was given by
Bandyopadhyay et al.~\cite{bandyopadhyay02}. We give a particularly
short proof by taking advantage of Weil sums.

\begin{theorem}\label{primepower}
Let\/ $\F_q$ be a finite field with odd characteristic $p$. 
Denote by $B_a  = \{ v_{a,b} 
\,|\, b\in \F_q\}$ the set of vectors given by 
$$v_{a,b}=q^{-1/2} \big(\omega_p^{\tr(ax^2+bx)}\big)_{x\in \F_q}.$$
The standard basis and the sets $B_a$, with $a\in \F_q$, 
form an extremal set of $q+1$ mutually unbiased bases of\/ $\C^q$.  
\end{theorem}
\begin{proof}
By definition
\begin{equation}\label{eq:sum}
 |\scal{v_{a,b}}{v_{c,d}}| = \Bigg| \frac{1}{q}\sum_{x\in \F_q} 
\omega_p^{\tr((c-a)x^2+(d-b)x)} \Bigg|.
\end{equation}
Suppose that $a=c$. The right hand side evaluates to 1 if $b=d$, and
to 0 if $b\neq d$. This proves that $B_a$ is an orthonormal basis.
The coefficients of the vector $v_{a,b}$ have
absolute value $q^{-1/2}$, hence $B_a$ is mutually unbiased with the
standard basis. On the other hand, 
if $a\neq c$, then the right hand side evaluates to 
$q^{-1/2}$ by
Lemma~\ref{weil}, which proves that the bases $B_a$ and $B_c$ are mutually unbiased.~\qed
\end{proof}

\begin{example}
In dimension~3, this construction yields 
the bases
$$ 
\begin{array}{rcccrrr}
B_0 & = & \{v_{0,0}, v_{0,1}, v_{0,2}\} 
&= \{& 3^{-1/2} (1,1,1), & 3^{-1/2}(1,\omega_3,\omega_3^2),&  3^{-1/2}(1,\omega_3^2,\omega_3)\},\\[1ex]
B_1 & = & \{v_{1,0}, v_{1,1}, v_{1,2}\}
&= \{ & 3^{-1/2} (1,\omega_3,\omega_3),  
&3^{-1/2}(1,\omega_3^2,1),  
&3^{-1/2}(1,1,\omega_3^2)\},\\[1ex]
B_2 & = & \{v_{2,0}, v_{2,1}, v_{2,2}\} 
&= \{ & 
3^{-1/2}(1,\omega_3^2,\omega_3^2),  
&3^{-1/2}(1,\omega_3,1),
&3^{-1/2}(1,1,\omega_3)
\},
\end{array}
$$
which form together with the standard basis four mutually unbiased bases. 
\end{example}

\section{Even Prime Powers} 
We showed in the last section that extremal sets of $q+1$ mutually
unbiased bases exist in dimension $q$ if $q$ is a power of an odd
prime. In this section we treat the case when $q$ is a power of
two. We cannot use Weil sums because Lemma~\ref{weil} does not apply
in even characteristics. However, it turns out that exponential sums
over a finite Galois ring can serve as a substitute.

\nix{
\bigskip
Galois rings $\GR(p^k, n)$ for $p$ prime, $k,n \in \N$ have been used
for sequence design in communications \cite{KHCH:96} and to show the
$\Z_4$-linearity of the Kerdock codes \cite{HKCSS:94,CCKS:97}. We use
the special case of the Galois rings $\GR(4,n)$ (see
\cite{HKCSS:94,KHC:95,YHKS:96,Carlet:2000}) to construct a set of MUBs
in dimension $d=2^n$. With the exception of the standard basis---which
without loss of generality can be assumed to be a member of any
family of MUBs---the bases we are going to construct have the feature
that they are defined over the set of amplitudes (up to a global
normalization factor) $\{ \pm 1, \pm i\}$, i.\,e., each vector has
entries over the complex fourth roots of unity. We recall some basic
facts about the Galois rings $\GR(4,n)$.
}
 
We recall some elementary facts about finite Galois rings,
see~\cite{Wan97}\nix{\cite{HKCSS:94,KHC:95,YHKS:96,Carlet:2000}} 
for more details.  Let
$\Z_4$ denote the residue class ring of integers modulo 4. Denote by
$\langle 2\rangle$ the ideal generated by $2$ in $\Z_4[x]$. A
monic polynomial $h(x)\in \Z_4[x]$ is called \textsl{basic
primitive}\/ if and only if its image in $\Z_4[x]/\langle
2\rangle\cong \Z_2[x]$ under the canonical map is a primitive
polynomial in $\Z_2[x]$.  Let $h(x)$ be a monic basic primitive
polynomial of degree $n$.  The ring $\GR(4,n) = \Z_4[x]/\langle
h(x)\rangle$ is called the Galois ring of degree $n$ over $\Z_4$.

The construction ensures that $\GR(4,n)$ has $4^n$ elements.  The
element $\xi = x+\langle h(x)\rangle$ is of order $2^n-1$.  Any
element $r\in\GR(4,n)$ can be uniquely written in the form $ r=a+2b$,
where $a,b\in {\cal T}_n = \{ 0, 1, \xi, \ldots, \xi^{2^n-2} \}.$ This
representation in terms of the Teichm{\"u}ller set ${\cal T}_n$ is
convenient, since it allows us to characterize the units of $\GR(4,n)$ as
the elements $a+2b$ with $a\neq 0$.

The automorphism $\sigma\colon \GR(4,n)\rightarrow \GR(4,n)$ defined
by $\sigma(a+2b)=a^2+2b^2$ is called the Frobenius automorphism.  This
map leaves the elements of the prime ring $\Z_4$ fixed. All automorphisms of
$\GR(4,n)$ are of the form $\sigma^k$ for some integer $k\ge 0$.  The
trace map $\tr\colon \GR(4,n)\rightarrow \Z_4$ is defined by $\tr(x) =
\sum_{k=0}^{n-1} \sigma^k(x)$. 

\begin{lemma}\label{charsums} 
Keep the notation as above. 
The exponential sum $\Gamma\colon \GR(4,n)\rightarrow \C$ defined by 
$\Gamma(r) = \sum_{x \in {\cal T}_n}
\exp(\frac{2\pi i}{4} \; {\rm tr}(rx))$ satisfies 
\[
|\Gamma(r)| = \left\{
\begin{array}{c@{\quad}l}
0 & \; \text{\rm if} \; r\in 2 {\cal T}_n, \; r \not=0, \\
2^n & \; \text{\rm if} \; r=0,\\
\sqrt{2^n} & \; \text{\rm otherwise}.
\end{array}
\right.
\]
\end{lemma}
The above lemma is proved in~\cite[Lemma 3]{Carlet:2000}, see also
\cite{YHKS:96}. This lemma will be crucial in the next construction
of mutually unbiased bases.

\begin{theorem}
Let\/ $\GR(4,n)$ be a finite Galois ring with Teichm\"uller set ${\cal T}_n$. 
For  $a\in {\cal T}_n$, denote by $M_a = \{ v_{a,b}\,|\, b\in {\cal T}_n \}$
the set of vectors given by 
$$v_{a,b}=2^{-n/2} \left(\exp\left(\frac{2\pi i}{4} {\tr(a+2b)x}\right)\right)_{x\in {\cal T}_n}.$$%
The standard basis and the sets $M_a$, with $a\in {\cal T}_n$, 
form an extremal set of $2^n+1$ mutually unbiased bases of\/ $\C^{2^n}$.  
\end{theorem}
\begin{proof}
By definition, 
$$ |\scal{v_{a,b}}{v_{a',b'}}| = \frac{1}{2^n}\left| \sum_{x \in {\cal T}_n}
\exp \left(\frac{2 \pi i}{4} \,\tr\big((a'-a) + 2(b'-b)\big)x \right) \right| 
$$ 
If both vectors belong to the same basis, \textit{i.e.}, when $a=a'$, then
Lemma~\ref{charsums} shows that the right hand side evaluates to 0
in case $b\neq b'$, and to 1 in case $b=b'$. This shows that $M_a$ is an
orthonormal basis. 

If the vectors belong to different bases, \textit{i.e.}, when $a\neq a'$, then 
Lemma~\ref{charsums} shows that $|\scal{v_{a,b}}{v_{a',b'}}|=2^{-n/2}$,
hence $M_a$ and $M_{a'}$ are mutually unbiased.
The entries of the vectors $v_{a,b}$ have absolute
value $2^{-n/2}$, thus the standard basis and $M_a$ are mutually
unbiased for all $a\in \GR(4,n)$.~\qed 
\end{proof}

\begin{example}
We illustrate this construction by deriving five mutually unbiased
bases in $\C^4$. In this case, the Galois ring $\GR(4,2)=
\Z_4[x]/\langle x^2+x+1\rangle$ with 16 elements is the basis of the
construction. The Teichm{\"u}ller set is given by ${\cal T}_2 = \{ 0,
1, 3\xi + 3, \xi\}$. Recall that an element of $\GR(4,2)$ can be
represented in the form $a+2b$ with $a,b\in {\cal T}_2$. By
definition, $\tr(a+2b)=a+2b+a^2+2b^2$. Computing the basis vectors
yields
$$ 
\newcommand{\ps}{\phantom{..}}
\newcommand{\mps}{\phantom{.}}
\begin{array}{lcl@{}llll}
M_0 &=& \big\{&\frac{1}{2}(1,\ps 1,\ps 1,\ps 1), 
&\frac{1}{2}(1,1,-1,-1), 
&\frac{1}{2}(1,-1,-1,1), 
&\frac{1}{2}(1,-1,1,-1) \big\},\\[1ex]
M_1 &=& \big\{&\frac{1}{2}(1,-1,-i,-i),
&\frac{1}{2}(1,-1,\ps i,\ps i),
&\frac{1}{2}(1,\ps 1,\ps i,-i),
&\frac{1}{2}(1,\ps 1,-i,\ps i) \big\},\\[1ex]
M_{3\xi+3} &=&  \big\{&\frac{1}{2}(1,-i,-i,-1),
&\frac{1}{2}(1,-i,\ps i,\ps 1),
&\frac{1}{2}(1,\ps i,\ps i,-1),
&\frac{1}{2}(1,\ps i,-i,\ps 1) \big\},\\[1ex]
M_{\xi} &=&  \big\{&\frac{1}{2}(1,-i,-i,-1),
&\frac{1}{2}(1,-i,\ps i,\ps 1),
&\frac{1}{2}(1,\ps i,\ps i,-1),
&\frac{1}{2}(1,\ps i,-i,\ps 1) \big\}.
\end{array}
$$ 
These four bases and the standard basis form an extremal set of five
mutually unbiased bases of $\C^4$. 
\end{example}

\section{Non Prime Powers}\label{general}
In the previous two sections, we established that the number $N(d)$ of
mutually unbiased bases in dimension $d$ attains the maximal possible
value, $N(d)=d+1$, when $d$ is a prime power.  In contrast, the exact
value of $N(d)$ is not known for any dimension $d$ which is divisible
by at least two distinct primes, not even in small dimensions such as $d=6$. 

The problem to determine $N(d)$ is similar to the combinatorial
problem to determine the number $M(d)$ of mutually orthogonal Latin
squares of size $d\times d$. The number $M(d)$ is exactly known for
prime powers but not in general when $d$ is divisible by at least two
distinct primes, see~\cite{beth99,laywine98} for more details.  Lower
bounds on the number of mutually orthogonal Latin squares can be
obtained with the help of a lemma by MacNeish. Our next result formulates a similar statement for the number $N(d)$ of mutually unbiased bases.

\begin{lemma}\label{macneish}
Let $d = p_1^{a_1} \cdots p_r^{a_r}$ be a factorization of $d$ into distinct 
primes $p_i$. Then $$N(d) \geq {\rm min} \, \{ N(p_1^{a_1}), N(p_2^{a_2}),
\dots, N(p_r^{a_r})\}.$$ 
\end{lemma}
\begin{proof}
We denote the minimum by $m = {\rm min}_i \,
N(p_i^{a_i})$. Choose $m$ mutually unbiased bases $B^{(i)}_1,
\ldots, B^{(i)}_m$ of $\C^{p^{a_i}}$, for all $i$ in the range $1\le i\le r$. 
Then 
\[
\{ 
B^{(1)}_k \otimes \ldots \otimes B^{(r)}_k : k = 1, \ldots, m
\}
\]
is a set of $m$ mutually unbiased bases of $\C^d$. 
\qed
\end{proof}

An easily memorable form of the above lemma is $N(nm)\ge\min\{
N(n),N(m)\}$ for all $m,n\ge 2$.  A simple consequence is that
$N(d)\ge 3$ for all dimensions $d\ge 2$, that is, in each dimension
there are at least three mutually unbiased bases.

Many researchers in the quantum physics community seem to be under the
impression that $N(d)=d+1$ for all integers $d\ge 2$. However, there
is some numerical evidence that considerably fewer mutually unbiased
bases might be possible if the dimension is not a prime power. In
fact, a conjecture by Zauner on the existence of affine quantum
designs implies that $N(6)=3$ rather than $N(6)=7$, see~\cite{zauner99}. 

\begin{conjecture}[Zauner]
The number of mutually unbiased bases in dimension 6 is given by $N(6)=3$. 
\end{conjecture}
Apparently, Zauner did considerable numerical computations to bolster his
conjecture. Our computational experiments indicate 
that $N(d)$ is in general smaller than $d+1$ when $d$ is not a prime
power. 
\begin{problem}
Does $N(d)=d+1$ hold for any dimension $d\ge 2$ that is not a prime power? 
\end{problem}

Another interesting problem concerns lower bounds on $N(d)$. Recall
that for mutually orthogonal Latin squares, $M(d)\rightarrow \infty$
for $d\rightarrow \infty$, as shown by Chowla, Erd\H{o}s, and
Strauss~\cite{chowla60}. It is natural to ask whether a similar
property holds for the number of mutually unbiased bases:
\begin{problem}
Does $N(d)\rightarrow \infty$ for $d\rightarrow \infty$ hold? 
\end{problem}
More constructions of mutually unbiased bases are needed to prove such
a result. A result similar to Wilson's theorem on the number
of mutually orthogonal Latin squares~\cite{wilson74} would be
particularly interesting.

%%%%%%%%%%%%%%%%%%%%%%%%%%%%%%%%%%%%%%%%%%%%%%%%%%%%%%%%%%%%
%
% Section: Conclusions and Outlook
%
%%%%%%%%%%%%%%%%%%%%%%%%%%%%%%%%%%%%%%%%%%%%%%%%%%%%%%%%%%%%

\section{Conclusions}
Mutually unbiased bases are basic primitives in quantum information
theory. They have applications in quantum cryptography and the design
of optimal measurements. It is known that in dimension $d$ at most
$d+1$ mutually unbiased bases can exist. In this paper, we gave a
simplified proof of the fact that $d+1$ mutually unbiased bases exist
in $\C^d$ when $d$ is a prime power. 

Specifically, we were able to generalize the construction by Alltop to
powers of a prime $p\ge 5$.  Elementary estimates of Weil sums
allowed us to derive a particularly short proof of a theorem by
Wootters and Fields.  For dimensions $d=2^n$, we took advantage of
known properties of exponential sums over $\GR(4,n)$ to obtain
extremal sets of mutually unbiased bases.

An open problem is to determine the maximal number of mutually
unbiased bases when the dimension is not a prime power.  We derived an
elementary lower bound for abritrary dimensions and discussed some
conjectures and open problems. Finally, we recommend the mean
king's problem~\cite{vaidman87,englert01,aravind03} as an enjoyable
application of mutually unbiased bases.

\paragraph{Acknowledgments.}  
This research was supported by NSF grant EIA 0218582, Texas A\&M TITF,
as well as ARDA, CFI, ORDFC, MITACS, and NSA.  We thank Hilary
Carteret, Chris Godsil, Bruce Richmond, and Igor Shparlinski for
useful discussions.

\end{document}